\RequirePackage{ifpdf}
\ifpdf 
\documentclass[pdftex]{sigma}
\else
\documentclass{sigma}
\fi

\begin{document}

\renewcommand{\PaperNumber}{092}

\renewcommand{\thefootnote}{$\star$}

\FirstPageHeading

\ShortArticleName{Miscellaneous Applications of Quons}

\ArticleName{Miscellaneous Applications of Quons\footnote{This paper is a
contribution to the Proceedings of the 3-rd Microconference
``Analytic and Algebraic Me\-thods~III''. The full collection is
available at
\href{http://www.emis.de/journals/SIGMA/Prague2007.html}{http://www.emis.de/journals/SIGMA/Prague2007.html}}}

\Author{Maurice R. KIBLER}
\AuthorNameForHeading{M.R. Kibler}

\Address{Universit\'e de Lyon, Institut de Physique Nucl\'eaire,
Universit\'e Lyon 1 and CNRS/IN2P3,\\ 43 bd du 11 novembre 1918,
F-69622 Villeurbanne Cedex, France}
\Email{\href{mailto:m.kibler@ipnl.in2p3.fr}{m.kibler@ipnl.in2p3.fr}}

\ArticleDates{Received July 23, 2007, in f\/inal form September 21,
2007; Published online September 24, 2007}

\Abstract{This paper deals with quon algebras or deformed oscillator algebras,
for which the deformation parameter is a root of unity. We show the interest
of such algebras for fractional supersymmetric quantum mechanics, angular
momentum theory and quantum information. More precisely, quon algebras are
used for (i) a realization of a generalized Weyl--Heisenberg algebra from
which it is possible to associate a fractional supersymmetric dynamical system,
(ii) a polar decomposition of SU$_2$ and (iii) a construction of mutually
unbiased bases in Hilbert spaces of prime dimension. We also brief\/ly discuss
(symmetric informationally complete)
positive operator valued measures in the spirit of (iii).}


\Classification{81R50; 81R05; 81R10; 81R15}

\section{Introduction}

Deformed oscillator algebras are spanned by generalized bosons
or fermions which are often referred to as quons, $q$-bosons,
parafermions (parabosons) or $k$-fermions
\cite{ArikCoon, Kury, Jannus, Bieden, Sun, Mac, Chaturandco, Gren,
DaoHasKib-1, DaoHasKib-2}. Such objects are of paramount importance
for generating $q$-deformed Lie algebras. In this respect, quons
can be used to derive $q$-deformed Lie (super)algebras associated
with $q$-deformed models in atomic, molecular and condensed matter
physics as well as in nuclear and particle physics.

In recent years, the use of deformed oscillator algebras proved to be useful
for many applications in quantum mechanics. For instance, one- and two-parameter
deformations of oscillator algebras and Lie algebras were applied to
intermediate statistics
\cite{GeSu, Mart, LeeYu, SuGe, TRMK, Vladimir,
Hsu, Granov, Chai, Gupta, Monteiro, Gong, DaoKib1995}
and to molecular and nuclear physics
\cite{Witten90, Iwao, Bonat1, Chang1, Chang2, Bonat2,
Bonat3, Bonat4, Jenko, Barbier}.
Along this vein, it was shown recently that $q$-bosons
play a central role for quantum integrable systems \cite{Kundu}.

It is the aim of the present work to show the interest of quons,
when $q$ is a $k$th primitive root of unity, for fractional supersymmetric
quantum mechanics of order $k$ and for the determination of mutually
unbiased bases used in quantum information. This paper takes it origin
in an invited talk at the $3^{\rm rd}$
International Microconference ``Analytic and Algebraic Methods
in Physics''. It presents a review character although it exhibits
some original aspects partly presented in schools and conferences.

The organisation of the paper is as follows. Section~\ref{section2} is devoted to some
generalities on quon algebras and $k$-fermions which are objects interpolating
between bosons and fermions. Section~\ref{section3} deals with fractional supersymmetric
quantum mechanics of order $k$. In Section~\ref{section4} we show how to construct the
Lie algebra of SU$_2$ from two quon algebras. This leads to a polar
decomposition which is used in Section~\ref{section5} for generating mutually unbiased
bases (MUBs) in $d$-dimensional Hilbert spaces with $d$ prime.  The approach
followed for MUBs is also suggested for deriving certain positive operator
valued measures (POVMs) in f\/inite-dimensional Hilbert spaces.

Regarding the notations, let us mention that
$\delta _{a,b}$ stands for the Kronecker symbol for $a$ and
$b$, the bar indicates complex conjugation, $A^{\dagger}$
denotes the adjoint of the operator $A$, $\mathbb{I}$ is
the identity operator, and $[A,B]_q := AB - qAB$ so that
$[A,B]_1$ (respectively $[A,B]_{-1}$) is the commutator $[A,B]$
(respectively anticommutator $\{A,B\}$) of the operators $A$ and $B$.

\section[Quons and $k$-fermions]{Quons and $\boldsymbol{k}$-fermions}\label{section2}

We f\/irst def\/ine a quon algebra or $q$-deformed oscillator algebra for $q$ a
root of unity.

\begin{definition}\label{definition1}
The three operators $a_-$, $a_+$ and $N_a$ such that
              \begin{gather}
 [ a_- , a_+ ]_q = \mathbb{I}, \qquad
 [ N_a , a_{\pm} ] = {\pm} a_{\pm}, \qquad
 a_{\pm}^k = 0, \qquad N_a ^{\dagger} = N_a,
              \label{Aq(x)}
              \end{gather}
where
              \begin{gather}
q := \exp({2\pi { i }}/{k}), \quad k \in \mathbb{N} \setminus \{ 0,1 \},
              \label{def(q,k)}
              \end{gather}
def\/ine a quon algebra or $q$-deformed oscillator algebra denoted as
$A_q(a_-, a_+, N_a)$. The opera\-tors $a_-$ and $a_+$ are referred to
as quon operators. The operators $a_-$, $a_+$ and $N_a$ are called
annihilation, creation and number operators, respectively.
\end{definition}

Def\/inition \ref{definition1} dif\/fers from the one by Arik and Coon \cite{ArikCoon}
in the sense that we take $q \in S^1$ instead of $0 < q < 1$. For $k=2$
(respectively $k \to \infty$), the quon operators coincide with the ordinary
fermion (respectively boson) operators.

For arbitrary $k$, the quon operators
$a_-$ and $a_+$ are not connected via Hermitian conjugation. It is only for
$k=2$ or $k \to \infty$ that we may take $a_+ = a_-^{\dagger}$. In general
({\it i.e.}, for $k \not= 2$ or $k \not\to \infty$), we have
$a_{\pm}^{\dagger} \not= a_{\mp}$. Therefore, it is natural to consider the
so-called $k$-fermionic algebra~$\Sigma_q$ with the generators
$a_-$,
$a_+$,
$a_+^+ := a_+^{\dagger}$,
$a_-^+ := a_-^{\dagger}$, and $N_a$ \cite{DaoHasKib-1, DaoHasKib-2}. The def\/ining
relations for~$\Sigma_q$ correspond to the ones of
$A_q       (a_-,   a_+,   N_a)$ and
$A_{\bar q}(a_+^+, a_-^+, N_a)$ complemented by the relation
              \[
 a_- a_+^{+} - q^{- \frac{1}{2}} a_+^{+} a_- = 0 \quad \Leftrightarrow \quad
 a_+ a_-^{+} - q^{  \frac{1}{2}} a_-^{+} a_+ = 0.
              \]
Observe that for $k = 2$ or $k \to \infty$, the latter relation
corresponds to an identity. The operators~$a_-$, $a_+$, $a_+^+$ and $a_-^+$
are called $k$-fermion operators and we also use the terminology
$k$-fermions in analogy with fermions and bosons
\cite{DaoHasKib-1}. They clearly interpolate between fermions
and bosons.

In a way similar to the one used for ordinary fermions and ordinary bosons,
we can def\/ine the coherence factor $g^{(m)}$ for an assembly of $m$
$k$-fermions. Such a def\/inition reads
              \[
g^{(m)} := { \langle \left( a_-^+ \right)^m
                     \left( a_-   \right)^m \rangle \over
             \langle a_-^+ a_- \rangle^m },
          \label{g(m)}
              \]
where
              \[
\langle X \rangle := { (z | X | z) \over (z | z) },
          \label{meanvalue}
              \]
stands for the average value of the operator $X$ on the $k$-fermionic
coherent state $|z)$ def\/ined in~\cite{DaoHasKib-2}. A simple calculation in
$\Sigma_q$ shows that
              \begin{gather}
 \big| g^{(m)} \big| =\left\{\begin{array}{ll}  0 & {\rm for} \ m >   k-1,\\
                       1 & {\rm for} \ m \leq k-1.
                       \end{array}\right.
              \label{0ou1}
              \end{gather}

From equation (\ref{0ou1}), we see once again that $k=2$ corresponds to
ordinary
fermions and $k \to \infty$ to ordinary bosons. Equation (\ref{0ou1}) is in
agreement with a generalized Pauli exclusion principle according to which,
for a many-particle system, a $k$-fermionic state (corresponding to a spin~$1/k$) cannot be occupied by more than $k-1$ identical $k$-fermions
\cite{DaoHasKib-1}.

To close this section, let us mention that the $k$-fermions introduced in
\cite{DaoHasKib-1, DaoHasKib-2} share some common properties with the
parafermions of order $k-1$ discussed
in \cite{RubSpi, BeckersDebergh, Debergh, Khare, Filippov, Durand, Klishevich}.
Indeed, a parafermionic algebra of order $k-1$ corresponds to a fractional
supersymmetric algebra of order $k$.

\section{Quons and fractional supersymmetry}\label{section3}

\subsection{Fractional supersymmetric system}\label{section3.1}

Following Rubakov and Spiridonov \cite{RubSpi},
who initially considered the case
$k=3$, we start with the def\/inition of a fractional supersymmetric system
of order $k$ \cite{DaoKib04}.

\begin{definition}\label{definition2}
For f\/ixed $k$ in $\mathbb{N} \setminus \{ 0,1 \}$, a fractional supersymmetric
quantum system of order~$k$, or $k$-SUSY system in short, is a doublet
$(H, Q)_k$ of linear operators $H$ and $Q$, acting on a~separable Hilbert
space ${\cal H}$, such that $H$ is self-adjoint and
              \[
Q_- := Q, \qquad
Q_+ := Q^{\dagger}, \qquad
Q_{\pm}^k = 0, \qquad
\sum_{s=0}^{k-1} Q_-^{k-1-s}Q_+Q_-^s = Q_-^{k-2}H, \qquad
[H , Q_{\pm}] = 0.
          \label{HQQ_k}
              \]
The operators $H$ and $Q_{\pm}$ are called the Hamiltonian and the supercharges
of the $k$-SUSY system.
\end{definition}

By way of illustration, let us show that the case $k= 2$ corresponds to Witten's
approach of ordinary supersymmetric quantum mechanics. According to
Witten~\cite{Witten1},
a triplet of linear operators $(H, P, S)$ of linear operators
$H$, $P$ and $S$, with $P$ bounded, def\/ines a supersymmetric quantum
system if
              \[
S = S^{\dagger}, \qquad
H = S^2, \qquad
\{ S , P \} = 0, \qquad
P^2 = \mathbb{I}, \qquad
P = P^{\dagger}.
          \label{Wittenkegaldeux}
              \]
By putting
          \[
Q_{\pm} = \frac{1}{2} S (1 \pm P),
          \]
we get the relations
          \[
Q_+ = Q_-^{\dagger}, \qquad
Q_{\pm}^2 = 0, \qquad
\{ Q_- , Q_+ \} = H, \qquad
[H , Q_{\pm}] = 0, \qquad H = H ^{\dagger},
          \]
which correspond to a $(H, Q)_2$ system or ordinary supersymmetric
system with a $Z_2$-grading involving fermionic and
bosonic states.

Going back to the general case, the doublet $(H, Q)_k$,
with arbitrary $k$, def\/ines a
$k$-SUSY system, with a $Z_k$-grading, for which the Hamiltonian $H$
and the two (nonindependent) supercharges $Q_{\pm}$ are up to now formal
operators. A natural question arises: How to f\/ind realizations of $H$
and $Q_{\pm}$? In this respect, the def\/inition of a generalized
Weyl--Heisenberg algebra is essential~\cite{DaoKib04}.

\subsection[Generalized Weyl-Heisenberg algebra]{Generalized Weyl--Heisenberg algebra}\label{section3.2}

\begin{definition}\label{definition3}
Let $W_k(f)$, where $f = \{ f_s : s = 0, 1, \ldots, k-1\}$ is a set of $k$
functions, be the algebra spanned by the four linear operators
$X_+$, $X_-$, $N$ and $K$ acting on the space ${\cal H}$ and
satisfying
    \begin{gather}
X_+ = X_-^{\dagger}, \qquad
N   =   N^{\dagger}, \qquad
KK^{\dagger} = K^{\dagger}K = \mathbb{I}, \qquad
K^k = \mathbb{I}
\label{WH1}
    \end{gather}
and
    \begin{gather}
[X_- , X_+] = \sum_{s=0}^{k-1} f_s(N)\Pi_s, \qquad
\Pi_s = \frac{1}{k} \sum_{t=0}^{k-1} q^{-st} K^t, \qquad
\label{WH2}
\\
[N , X_{\pm}] = {\pm} X_{\pm}, \qquad
[K , X_{\pm}]_{q^{\pm 1}} = 0, \qquad
[K , N] = 0,
\label{WH3}
    \end{gather}
where $q := \exp({2\pi { i }}/{k})$ and $k \in \mathbb{N} \setminus \{ 0,1 \}$.
\end{definition}

It should be realized that, for f\/ixed $k$, equations
(\ref{WH1})--(\ref{WH3}) def\/ine indeed a family of
generalized Weyl--Heisenberg algebras $W_k(f)$. The
various members of the family are distinguished by
the various sets $f$.

It is possible to f\/ind a realization of $W_k(f)$ in terms of one pair
($f_- , f_+$) of $k$-fermions with
\[
[f_- , f_+]_q = \mathbb{I}, \qquad f_{\pm}^k = 0
\]
and $k$ pairs ($ b(s)_- , b(s)_+$) of generalized bosons with
\[
[b(s)_-, b(s)_+] = f_s(N), \qquad s = 0, 1, \ldots, k-1,
\]
such that any $k$-fermion operator commutes with any generalized boson
operator. Indeed, by introducing
\[
b_{\pm} = \sum_{s=0}^{k-1} b(s)_{\pm} \Pi_s
\]
one can take
\[
X_- = b_- \left( f_-   +    \frac{ f_+ ^{k-1}}{ [k-1] _q !} \right),      \qquad
X_+ = b_+ \left( f_-   +    \frac{ f_+ ^{k-1}}{ [k-1] _q !} \right)^{k-1}, \qquad
K = [ f_- , f_+],
\]
where
          \[
  \forall\, n \in \mathbb{N}^{\ast} :\quad  \left[ n \right]_q := \frac{1-q^n}{1-q}, \qquad
  \left[ n \right]_q! :=
  \left[ 1 \right]_q
  \left[ 2 \right]_q  \cdots
  \left[ n \right]_q.
          \]

\subsection[Realizations of $k$-SUSY systems]{Realizations of $\boldsymbol{k}$-SUSY systems}\label{section3.3}

As an important result, the following proposition shows that, for f\/ixed $k$
and given $f$, it is possible to associate a $k$-SUSY system, characterized
by a specif\/ic doublet $(H, Q)_k$, with the algebra $W_k(f)$.

\begin{proposition}\label{proposition1}
For a fixed value of $k$ and a given set $f$, the relations
\[
Q_- := Q         = X_- (1 - \Pi_1), \qquad
Q_+ := Q^\dagger = X_+ (1 - \Pi_0)
\]
and
\[
H = (k-1)X_+X_-
 - \sum_{s=3}^{k}  \sum_{t=2}^{s-1} (t-1) f_t(N-s+t) \Pi_s
 - \sum_{s=1}^{k-1}\sum_{t=s}^{k-1} (t-k) f_t(N-s+t) \Pi_s
\]
generate a $(H, Q)_k$ system associated with the generalized
Weyl--Heisenberg algebra $W_k(f)$. In addition, the Hamiltonian $H$ can be
decomposed as
\[
H = \sum_{s=0}^{k-1} {H}_{k-s} \Pi_{k-s},
\]
where the various operators ${H}_{k-s}$ are isospectral Hamiltonians. Each
Hamiltonian ${H}_{k-s}$ acts on a subspace ${\cal H}_{k-s}$ of the $Z_k$-graded
Hilbert space
\[
{\cal H} = \bigoplus_{s=0}^{k-1} {\cal H}_{k-s},
\]
with $H_k \equiv H_0$.
\end{proposition}

In the light of Proposition \ref{proposition1}, we foresee that a $k$-SUSY system
with a $Z_k$-grading can be
considered as a superposition of $k-1$ ordinary supersymmetric
subsystems (corresponding to $k=2$) with a $Z_2$-grading~\cite{DaoKib04}.

As an example, we consider the case of the $Z_3$-graded supersymmetric
oscillator corresponding to
\[
k=3 \ \ \Rightarrow \ \ q = \exp \left( {2 \pi {i} \over 3} \right)
\]
with
\[
f_s(N) = 1, \quad s = 0, 1, 2, \ \Rightarrow \ [X_- , X_+] = \mathbb{I}.
\]
The corresponding algebra $W_3(f)$ can be represented by
\begin{gather*}
X_- = b_- \left( f_-   +    \frac{ f_+ ^{2}}{ [2] _q !} \right),    \qquad
X_+ = b_+ \left( f_-   +    \frac{ f_+ ^{2}}{ [2] _q !} \right)^{2}, \\
K = f_- f_+   -   f_+ f_-, \qquad
N = b_+ b_-,
\end{gather*}
in terms of $3$-fermions ($f_-, f_+$) and ordinary bosons ($b_-, b_+$).
The system $(H, Q)_3$ associated with $W_3(f)$ is def\/ined by
\[
Q_- := Q           = X_- (\Pi_0 + \Pi_2), \qquad
Q_+ := Q^{\dagger} = X_+ (\Pi_1 + \Pi_2)
\]
and
\[
H = \left( 2 X_+ X_-  -  1 \right) \Pi_3 +
    \left( 2 X_+ X_-  +  1 \right) \Pi_2 +
    \left( 2 X_+ X_-  +  3 \right) \Pi_1,
\]
where
\begin{gather*}
\Pi_0 = \frac{1}{3} \left( 1 + q^3 K + q^3 K^2 \right), \qquad
\Pi_1 = \frac{1}{3} \left( 1 + q^1 K + q^2 K^2 \right), \\
\Pi_2 = \frac{1}{3} \left( 1 + q^2 K + q^1 K^2 \right),
\end{gather*}
with $\Pi_3 \equiv \Pi_0$. In terms of $3$-fermions and
ordinary bosons, we have
\[
Q_- = b_- f_+ \left( f_-^2 -  q f_+   \right), \qquad
Q_+ = b_+     \left( f_-   -  q f_+^2 \right) f_-
\]
and
\[
H = 2 b_+ b_-  -  1  +  2(1 - 2q) f_+ f_-  +  2(1 + 2 q) f_+ f_- f_+ f_-.
\]
Finally, the energy spectrum of $H$ reads
\[
{\rm spectrum} (H) = 1 \oplus 2 \oplus 3 \oplus 3 \oplus \cdots,
\]
a symbolic writing to mean that it contains equally spaced levels with
a nondegenerate ground state (denoted as 1), a doubly degenerate f\/irst
excited state (denoted as 2) and an inf\/inite sequence of triply
degenerate excited states (denoted as 3).

Other sets $f$ lead to other fractional supersymmetric dynamical systems
\cite{DaoKib06}. For instance, the case
\[
f_s(N) = a N + b, \qquad s = 0, 1, \ldots, k-1, \qquad a \in \mathbb{R}, \qquad b \in \mathbb{R}
\]
corresponds to translational shape-invariant systems as for example
the harmonic oscillator system (for $a=0$ and $b > 0$),
the Morse system (for $a < 0$ and $b > 0$) and the
P\"oschl--Teller system (for $a > 0$ and $b > 0$). Furthermore, the case
\[
f_s(N) = f_s, \qquad s = 0, 1, \ldots, k-1
\]
corresponds to cyclic shape-invariant systems like the Calogero--Vasiliev
system for $ k=2 $, $f_0 = 1+c$ and $f_1 = 1-c$ with $c \in \mathbb{R}$.

\section{Quons and polar decomposition}\label{section4}

The approach presented in this section is a review based on the original
developments given in \cite{Inde, CCCC, IJMP1, KibPla06, AlbKib07}. Here,
we shall limit ourselves to those aspects which are relevant for Section~5.

We start with two commuting quon algebras $A_q(a_-, a_+, N_a) \equiv A_q(a)$
with $a = x, y$ corresponding to the same value of the deformation parameter
$q$. Their
generators satisfy equations~(\ref{Aq(x)}) and
(\ref{def(q,k)}) with $a = x, y$ and $[X, Y]=0$
for any $X$ in $A_q(x)$ and
    any $Y$ in~$A_q(y)$. Then, let us look for Hilbertian representations
of~$A_q(x)$ and~$A_q(y)$ on $k$-dimensional Hilbert spaces
${\cal F}_x$ and~${\cal F}_y$ spanned by the orthonormal bases
$\{ | n_1 ) : n_1 = 0, 1, \ldots, k-1 \}$ and
$\{ | n_2 ) : n_2 = 0, 1, \ldots, k-1 \}$, respectively. We easily
obtain the representations def\/ined by
      \begin{gather*}
  x_+ |n_1) = |n_1 + 1),                    \qquad
  x_+ |k-1) = 0,                            \qquad
  x_- |n_1) = \left[ n_1   \right]_q |n_1-1),  \\
  x_- |0)   = 0,                            \qquad
  N_x |n_1) = n_1 |n_1)
      \end{gather*}
and
      \begin{gather*}
  y_+ |n_2) = \left[ n_2+1 \right]_q |n_2+1),  \qquad
  y_+ |k-1) = 0,                            \qquad
  y_- |n_2) = |n_2 - 1),                    \\
  y_- |0) = 0,                              \qquad
  N_y |n_2) = n_2 |n_2),
      \end{gather*}
for $A_q(x)$ and $A_q(y)$, respectively.

The cornerstone of this approach is to def\/ine the two linear operators
              \[
  h := {\sqrt {N_x \left( N_y + 1 \right) }}, \qquad v_{ra} := s_x s_y,
              \]
with
\begin{gather*}
   s_x = q^{ a (N_x + N_y) / 2 } x_{+} +
   {e} ^{ {i} \phi_r / 2 }  {1 \over
  \left[ k-1 \right]_q!} (x_{-})^{k-1},
  \\
   s_y = y_{-}  q^{- a (N_x - N_y) / 2 }+
   {e} ^{ {i} \phi_r / 2 }  {1 \over
  \left[ k-1 \right]_q!} (y_{+})^{k-1},
\end{gather*}
where
          \[
 a \in \mathbb{R}, \qquad \phi_r = \pi (k-1) r, \qquad r \in \mathbb{R}.
          \]
The operators $h$ and $v_{ra}$ act on the states
\[
| n_1 , n_2 ) = | n_1) \otimes | n_2 )
\]
of the $k^2$-dimensional space ${\cal F}_x \otimes {\cal F}_y$.

We now adapt the trick used by Schwinger in his approach to angular momentum via
a coupled pair of harmonic oscillators. This can be done by introducing two new
quantum numbers~$J$ and~$M$ def\/ined by
          \[
  J := {1 \over 2} \left( n_1+n_2 \right),  \qquad
  M := {1 \over 2} \left( n_1-n_2 \right) \quad \Rightarrow \quad
  |J M \rangle := |J + M , J-M) = |n_1 , n_2)
          \]
Note that
          \[
j := \frac{1}{2}(k-1)
          \]
is an admissible value for $J$. Then, let us consider the ($k$-dimensional)
subspace $\epsilon(j)$ of the ($k^2$-dimensional) space
${\cal F}_x \otimes {\cal F}_y$ spanned by the basis
          \[
S = \{ |j , m \rangle : m = -j, -j+1, \ldots, j \}.
          \]
We guess that $\epsilon(j)$ is a space of constant angular momentum $j$. As a
matter of fact, we can check that $\epsilon(j)$ is stable under $h$ and
$v_{ra}$. Moreover, by def\/ining the operators $j_{\pm}$ and $j_z$ through
              \[
  j_+ := h           v_{ra},  \qquad
  j_- := v_{ra}^{\dagger} h,  \qquad
  j_z := \frac{1}{2} ( h^2 - v_{ra}^{\dagger} h^2 v_{ra} ),
  \label{definition of the j's}
              \]
we obtain
              \[
  \left[ j_z,j_{\pm} \right] = \pm j_{\pm},  \qquad
  \left[ j_+,j_- \right] = 2j_z,
  \label{adL su2}
              \]
for any $r$ in $\mathbb{R}$ and any $a$ in $\mathbb{R}$. The latter
commutation relations correspond to the Lie algebra of SU$_2$.

We have here a polar decomposition of $j_{\pm}$. Thus, from two $q$-deformed
oscillator algebras
we obtained a polar decomposition of the nondeformed Lie algebra of SU$_2$.
In addition, the complete set of commuting operators $\{ j^2 , j_z \}$, where
$j^2$ is the Casimir of SU$_2$ can be replaced by another
complete set of commuting operators, namely $\{ j^2 , v_{ra} \}$,
for which we have the following result. It is to be noted that the operators
$z$, def\/ined through $z | j,m \rangle = q^{j-m} | j,m \rangle$, and $v_{ra}$ can
be used to generate the Pauli group ${\cal P}_{2j+1}$, a subgroup of order
$(2j+1)^3$ of the linear group GL($2j+1 , \mathbb{C}$) \cite{AlbKib07}.

\begin{proposition}\label{proposition2}
For fixed $a$, $r$ and $j$, the common orthonormalized
eigenvectors of the commu\-ting operators $j^2$
and $v_{ra}$ can be taken in the form
              \begin{gather}
|j \alpha ; r a \rangle = \frac{1}{\sqrt{2j+1}} \sum_{m = -j}^{j}
q^{(j + m)(j - m + 1)a / 2 - j m r + (j + m)\alpha} | j , m \rangle,
\qquad \alpha = 0, 1, \ldots, 2j,
\label{j alpha r a in terms of jm}
              \end{gather}
or alternatively
              \begin{gather}
| j \alpha ; r a \rangle \equiv | a \alpha \rangle
= \frac{1}{\sqrt{d}} \sum_{k = 0}^{d-1}
q^{(d - k - 1)(k + 1)a / 2 + j (k-j) r - (k+1)\alpha} | k \rangle,
\qquad \alpha = 0, 1, \ldots, 2j.
\label{j alpha r a in terms of jm bis}
              \end{gather}
after introducing $k := j-m$, $d := 2j+1$, and $| k \rangle := | j , m \rangle$.
\end{proposition}

In other words, for f\/ixed $a$, $r$ and $j$, the space $\epsilon(j)$  can be
spanned by the basis
\[
B_{ra} = \{ |j \alpha ; ra \rangle : \alpha = 0, 1, \ldots, 2j \}.
\]
The replacement of the spherical basis $S$, adapted to the chain
SO$_3$ $\supset$ SO$_2$ and to the set $\{ j^2 , j_z \}$,
by the basis $B_{ra}$, adapted to the chain
SO$_3$ $\supset$ $Z_{2j+1}$ and to the set $\{ j^2 , v_{ra} \}$,
leads to a new form of the Wigner--Racah algebra of SU$_2$ \cite{CCCC}. Note
that the notation in (\ref{j alpha r a in terms of jm bis}) is particularly
appropriate for quantum information theory (the qudits
$| 0 \rangle, | 1 \rangle, \ldots, | d-1 \rangle$
in the expansion (\ref{j alpha r a in terms of jm bis}) constitute the
computational or canonical basis of a $d$-dimensional Hilbert space).

We shall see in Section \ref{section5} how the parameter $a$ can be chosen in such
a way to generate MUBs.

\section{MUBs and POVMs}\label{section5}

\subsection{MUBs}\label{section5.1}

We are now turn to the application of Proposition \ref{proposition2} to the determination
of MUBs in a f\/inite-dimensional Hilbert space. Let us recall that two
orthonormalized bases
$\{ |a \alpha \rangle : \alpha = 0, 1, \ldots$, $d-1 \}$ and
$\{ |b \beta  \rangle : \beta  = 0, 1, \ldots, d-1 \}$
of the $d$-dimensional Hilbert space $\mathbb{C}^d$, with an inner product
denoted as $\langle \ | \ \rangle$, are said to be mutually unbiased if
$| \langle a \alpha | b \beta \rangle | = 1/\sqrt{d}$ for $a \not= b$. In
other words, we have
\[
\left| \langle a\alpha | b\beta \rangle \right| =
\delta _{\alpha ,\beta } \delta_{a,b} + \frac{1}{\sqrt{d}} (1-\delta _{a,b}).
\]
We know that the number of MUBs in $\mathbb{C}^d$ is lesser or equal to $d+1$
and that the limit $d+1$ is reached when $d$ is a power of a
prime \cite{Ivano, Tolar, Balian, Wootters-Fields, Barnum, Bandyo, Chaturvedi,
Pittinger, Klappenecker-1, Bengtsson}.

Let us suppose that $d$ is the power of a prime. Then, we identify
$\mathbb{C}^d$
with the space $\epsilon(j)$ corresponding to the angular momentum
$j = (d-1)/2$. We begin by introducing the operators
    \begin{gather}
\Pi _{a\alpha } := |a\alpha \rangle \langle a\alpha |, \qquad
a = 0, 1, \ldots, 2j+1, \qquad
\alpha = 0, 1, \ldots, 2j.
\label{projector_pi_a_alpha}
    \end{gather}
Each of the operators $\Pi _{a\alpha }$ acting on $\mathbb{C}^d$ can be
considered as a vector in $\mathbb{C}^{d^2}$ endowed with the Hilbert--Schmidt
inner product. Thus, $\Pi _{a\alpha }$ can be developed as
    \begin{gather}
\Pi _{a\alpha } = \sum_{m = -j}^j \sum_{m' = -j}^j g_{mm'} (a\alpha) E_{mm'},
\label{projector_pi_a_alpha_en_E}
    \end{gather}
where
    \begin{gather}
E_{mm'} = |j, m \rangle \langle j, m'|.
\label{E_en_jmm'}
    \end{gather}
The calculation of
${\rm Tr}\big( \Pi _{a\alpha }^{\dagger} \Pi _{b\beta }\big)$
leads to
\[
\sum_{m = -j}^j \sum_{m' = -j}^j \overline{g_{mm'} (a\alpha)} g_{mm'} (b\beta)=
\delta _{\alpha ,\beta } \delta_{a,b} + \frac{1}{d} (1-\delta _{a,b}).
\]
By def\/ining the vectors
\[
s(a\alpha ) := \left(
s_{1}(a\alpha ),s_{2}(a\alpha ),\ldots,s_{d^{2}}(a\alpha) \right)
\]
with
\[
s_{i}(a\alpha ) := g_{mm'} (a\alpha), \qquad i = (j+m)(2j+1) + j + m' + 1,
\]
we get
    \begin{gather}
s(a\alpha ) \cdot s(b\beta ) =
\delta _{\alpha ,\beta } \delta_{a,b} + \frac{1}{d} (1-\delta _{a,b}),
\label{produit scalaire ss}
    \end{gather}
where $s(a\alpha ) \cdot s(b\beta )$ stands for the usual inner product
$\sum_{i=1}^{d^2} \overline{s_{i}(a\alpha )} s_{i}(b\beta )$ in
$\mathbb{C}^{d^2}$. Note that the relation (\ref{produit scalaire ss}) is
independent of the basis chosen for developing the operators $\Pi _{a\alpha }$.

As a result, the determination of the $d(d+1)$ vectors
$|a \alpha \rangle$ in $\mathbb{C}^{d}$, for $d$ a power of a prime, amounts to
f\/ind $d(d+1)$ operators $\Pi _{a\alpha }$ acting on $\mathbb{C}^{d}$ or to
f\/ind $d(d+1)$ vectors $s(a \alpha)$ in $\mathbb{C}^{d^2}$ satisfying
(\ref{produit scalaire ss}).

We may now use Proposition \ref{proposition2} derived from the quonic approach of
Section~\ref{section4} to obtain the following result.

\begin{proposition}\label{proposition3}
For fixed $r$ and $j=(d-1)/2$ with $d$ prime, the $d^2$ vectors
              \[
 |a \alpha \rangle := |j \alpha ; r a \rangle,  \qquad
 a     = 0, 1, \ldots, d-1, \qquad
 \alpha = 0, 1, \ldots, d-1,
              \]
where $|j \alpha ; r a \rangle$ is given by \eqref{j alpha r a in terms of jm}
                                        or \eqref{j alpha r a in terms of jm bis}
with $q := \exp({2\pi { i }}/{d})$, together with the $d$ vectors
              \[
 |a \alpha \rangle := |j , m \rangle, \qquad a = d, \qquad \alpha = j+m, \qquad
 m = -j, -j+1, \ldots, j,
              \]
constitute a complete set of $d+1$ MUBs in $\mathbb{C}^{d}$.
\end{proposition}

\begin{proof}
From \eqref{projector_pi_a_alpha},
     \eqref{projector_pi_a_alpha_en_E} and
     \eqref{E_en_jmm'}, we get
              \[
g_{mm'} (a\alpha) = \langle j , m   | a \alpha \rangle
                  \langle a \alpha | j , m'  \rangle.
          \label{g_en_jmaalpha}
              \]
Let us f\/irst consider the case $a = 0, 1, \ldots, d-1$. Equation
(\ref{j alpha r a in terms of jm}) leads to
              \begin{gather}
\langle j , m | a \alpha \rangle = \frac{1}{\sqrt{2j+1}}
q^{(j + m)(j - m + 1)a / 2 - j m r + (j + m)\alpha}.
          \label{produit_scalaire_jmaalpha}
	      \end{gather}
The introduction of (\ref{produit_scalaire_jmaalpha}) in
\[
s(a\alpha ) \cdot s(b\beta ) =
\sum_{m = -j}^j \sum_{m' = -j}^j \overline{g_{mm'} (a\alpha)} g_{mm'} (b\beta)
\]
yields
    \begin{gather}
s(a\alpha ) \cdot s(b\beta ) = \frac{1}{(2j+1)^2}
\sum_{m = -j}^j \sum_{m' = -j}^j
q^{(m - m')[(m + m' + 1)(a - b)/2 -(\alpha - \beta)]}
    \label{s point s en mm'}
    \end{gather}
for $a, b = 0, 1, \ldots, d-1$. By putting $k = j+m$ and $\ell = j+m'$,
equation (\ref{s point s en mm'}) can be rewritten~as
    \[
s(a\alpha ) \cdot s(b\beta ) =
\frac{1}{d^2} \sum_{k=0}^{d-1} \sum_{\ell = 0}^{d-1}
e^{i \pi (k-\ell) [(k + \ell - d) (a - b) - 2 (\alpha - \beta)]/d}.
    \]
In terms of the generalized quadratic Gauss sums \cite{Berndt}
              \[
S(u, v, d) := \sum_{k=0}^{d -1}  {e} ^{ {i} \pi (uk^2 + vk) / d },
   \label{generalized quadratic Gauss sum}
              \]
we obtain
    \[
s(a\alpha ) \cdot s(b\beta ) = \frac{1}{d^2} | S(u, v, d) |^2, \qquad
u = a - b, \qquad
v = 2 (\beta - \alpha) + d (b-a).
    \]
The calculation of $S(u, v, d)$, see \cite{AlbKib07} and \cite{Berndt}, leads to
    \[
s(a\alpha ) \cdot s(b\beta ) =
\delta _{\alpha ,\beta } \delta_{a,b} + \frac{1}{d} (1-\delta _{a,b}), \qquad
a, b  = 0, 1, \ldots, d-1, \qquad
\alpha = 0, 1, \ldots, d-1,
    \]
which proves that the $d$ bases
$B_{ra} := \{ | a \alpha \rangle : \alpha = 0, 1, \ldots, d-1 \}$
with $a = 0, 1, \ldots, d-1$ are mutually
unbiased. It is clear from (\ref{produit_scalaire_jmaalpha}) that the
bases $B_{ra}$ with
$a = 0, 1, \ldots, d-1$ and the basis
$B_d := \{  | d \alpha \rangle : \alpha = 0, 1, \ldots, d-1 \}$
are mutually unbiased. This completes the proof of Proposition \ref{proposition3}.
\end{proof}

\subsection{POVMs}\label{section5.2}

An approach similar to the one developed for MUBs can be set up
for symmetric informationally complete (SIC) POVMs
\cite{Zauner, Caves, Renes, Appleby, Grassl, Klappenecker-a2,
Klappenecker-2, Weigert}. We shall brief\/ly
discuss the starting point for a study of SIC-POVMs along the lines
of Section \ref{section5.1} (see \cite{AlbKib2} for more details).

A SIC-POVM in dimension $d$ can be def\/ined as a set of $d^2$ nonnegative
operators $P_x$ acting on $\mathbb{C}^{d}$ and satisfying
    \begin{gather}
{\rm Tr}\left( P_{x}P_{y}\right) =\frac{1}{d+1} (d \delta_{x,y} + 1), \qquad
\frac{1}{d}\sum_{x=1}^{d^{2}}P_{x}=\mathbb{I}, \qquad
P_{x}=|\Phi _{x}\rangle \langle \Phi _{x}|.
    \label{definition des P}
    \end{gather}
Let
\[
P_{x} = \sum_{m = -j}^j \sum_{m' = -j}^j f_{mm'} (a\alpha) E_{mm'}
\]
be the development of $P_{x}$ in terms of the operators $E_{mm'}$ and
let
\[
r(x) := \left( r_{1}(x), r_{2}(x), \ldots, r_{d^{2}}(x) \right)
\]
be the vector in $\mathbb{C}^{d^2}$ of components
\[
r_{i}(x) := f_{mm'} (a\alpha), \qquad i = (j+m)(2j+1) + j + m' + 1,
\]
It is immediate to show that
     \begin{gather}
r(x) \cdot r(y) = \frac{1}{d+1} (d \delta_{x,y} + 1),
     \label{produit scalaire rr}
     \end{gather}
a relation (independent of the basis chosen for developing $P_x$)
to be compared with (\ref{produit scalaire ss}).

The determination of the $d^2$ operators $P_x$ satisfying
(\ref{definition des P})
amounts to f\/ind $d^2$ vectors $|\Phi _{x} \rangle$ in~$\mathbb{C}^{d}$ or $d^2$ vectors $r(x)$ in~$\mathbb{C}^{d^2}$
satisfying (\ref{produit scalaire rr}). The search for solutions
of (\ref{produit scalaire rr}) is presently under progress.

\section{Open questions}\label{section6}

To close this paper, we would like to address a few questions which
arose during the conference. Possible future developments concern
(i) a f\/ield theory approach to $k$-fermions and their relation to
anyons, (ii) the classif\/ication, in terms of the sets $f$, of the
Weyl--Heisenberg algebras $W_k(f)$ that lead to integrable systems,
(iii) the passage from fractional supersymmetric quantum mechanics
to fractional supersymmetric nonHermitian quantum mechanics, as for
example along the lines of PT-symmetric regularizations \cite{Miloslav},
(iv) the Wigner--Racah algebra of the group SU$_2$ in the
$\{ j^2 , v_{ra}\}$ scheme, and (v) the construction of MUBs and
SIC-POVMs in an unif\/ied way.

Some comments regarding points (iv) and (v) are in order.

The $\{j^2 , v_{ra}\}$ scheme described in Section \ref{section4} constitutes an
alternative to the familiar $\{j^2 , j_z\}$ scheme of angular momentum
theory. As a further step, it would be interesting to f\/ind dif\/ferential
realizations of the operator $v_{ra}$ as well as realizations of the bases
$B_{ra}$ on the sphere $S^2$ for $j$ integer and on Fock--Bargmann spaces
(in 1 and 2 dimensions) for $j$ integer or half of an odd integer.

The preliminary study reported in Section \ref{section5} for MUBs and POVMs is based
on the replacement of
\begin{gather}
\left| \langle a\alpha | b\beta \rangle \right| =
\delta _{\alpha ,\beta } \delta_{a,b} + \frac{1}{\sqrt{d}} (1-\delta _{a,b}),
\label{biil1}
\\
\left| \langle \Phi_x | \Phi_y \rangle \right| =
\frac{1}{\sqrt{d+1}} {\sqrt{d \delta_{x,y} + 1}},
\label{biil2}
\end{gather}
corresponding to inner products in $\mathbb{C}^d$, by
\begin{gather}
s(a\alpha ) \cdot s(b\beta ) =
\delta _{\alpha ,\beta } \delta_{a,b} + \frac{1}{d} (1-\delta _{a,b}),
\label{bool1}
\\
r(x) \cdot r(y) =
\frac{1}{d+1} \left( d \delta_{x,y} + 1 \right),
\label{bool2}
\end{gather}
corresponding to inner products in $\mathbb{C}^{d^2}$,
respectively. We may ask
the question: Is it easier to f\/ind solutions of (\ref{biil1}), (\ref{biil2}) than
to f\/ind solutions of (\ref{bool1}), (\ref{bool2})? We do not have any answer. In
the case of MUBs, we have a general solution (see Proposition \ref{proposition3}) of
(\ref{bool1}) and thus (\ref{biil1}) for~$d$ prime
in the framework of angular momentum
theory. It would be interesting to extend this result for~$d$ a power of a
prime. In the case of SIC-POVMs, to prove or disprove the conjecture
according to which SIC-POMVs exist in any dimension amounts to prove
or disprove that (\ref{bool2}) has a solution in any dimension.

\medskip

\noindent
{\bf Note.}
After the submission (July 23,  2007) of the present paper for publication in
\textit{SIGMA}, the author became aware of a preprint dealing with the existence
of SIC-POVMs posted (July 20,  2007) on arXiv \cite{HallRao}. The main result in
\cite{HallRao} is that SIC-POVMs exist in all dimensions. As a~corollary of this
result, equation~(\ref{bool2}) admits solutions in any dimension.

\subsection*{Acknowledgements}

Some parts of the material reported here were worked out in collaboration
with Mohammed Daoud, Olivier Albouy, and Michel Planat. The present paper
is a contribution to the $3^{\rm rd}$ International
Microconference ``Analytic and Algebraic Methods in Physics'' (June 2007, Villa
Lanna, Prague); the author is very indebted to Miloslav Znojil
for organizing the conference and for useful comments; thanks are due
to Uwe G\"unther, Stefan Rauch-Wojciechowski, Artur Sergyeyev, Petr
\v{S}ulcp, and Pierguilio Tempesta for interesting discussions. This work
was also presented at the workshop ``Finite Projective Geometries in Quantum
Theory'' (August 2007, Astronomical Institute, Tatransk\'a Lomnica); the author
acknowledges the organizer, Metod Saniga, and the other participants for
fruitful interactions.

\pdfbookmark[1]{References}{ref}
\LastPageEnding

\end{document}